\documentclass[%
 aip,
 amsmath,amssymb,
 reprint,%
]{revtex4-1}

\usepackage{graphicx}
\usepackage{dcolumn}
\usepackage{bm}

\usepackage[utf8]{inputenc}
\usepackage[T1]{fontenc}
\usepackage{mathptmx}
\usepackage{upgreek} 
\usepackage{xcolor}
\usepackage[colorlinks=false, linkbordercolor=red, citebordercolor=green, urlbordercolor=cyan, pdfborderstyle={/S/U/W 1}]{hyperref}

\begin{document}

\preprint{AIP/123-QED}

\title{A room-temperature ion trapping apparatus with hydrogen partial pressure below $10^{-11}$~mBar}

\author{P.~Ob\v{s}il}
\affiliation{Department of Optics, Palack\'{y} University, 17. listopadu 12, 771 46 Olomouc, Czech Republic}
\author{A.~Le\v{s}und\'{a}k}
\affiliation{Institute of Scientific Instruments of the Czech Academy of Sciences, Kr\'{a}lovopolsk\'{a} 147, 612 64 Brno, Czech Republic}
\author{T.~Pham}
\affiliation{Institute of Scientific Instruments of the Czech Academy of Sciences, Kr\'{a}lovopolsk\'{a} 147, 612 64 Brno, Czech Republic}
\author{K. Lakhmanskiy}
\affiliation{Institut f\"{u}r Experimentalphysik, Universit\"{a}t Innsbruck, Technikerstra\ss e 25, 6020 Innsbruck, Austria}
\author{L.~Podhora}
\affiliation{Department of Optics, Palack\'{y} University, 17. listopadu 12, 771 46 Olomouc, Czech Republic}
\author{M.~Oral}
\affiliation{Institute of Scientific Instruments of the Czech Academy of Sciences, Kr\'{a}lovopolsk\'{a} 147, 612 64 Brno, Czech Republic}
\author{O.~\v{C}\'{i}p}
\affiliation{Institute of Scientific Instruments of the Czech Academy of Sciences, Kr\'{a}lovopolsk\'{a} 147, 612 64 Brno, Czech Republic}
\author{L.~Slodi\v{c}ka}
\email{slodicka@optics.upol.cz}
\affiliation{Department of Optics, Palack\'{y} University, 17. listopadu 12, 771 46 Olomouc, Czech Republic}

\date{\today}

\begin{abstract}
The lifetime of trapped ion ensembles corresponds to a crucial parameter determining the potential scalability of their prospective applications and is often limited by the achievable vacuum level in the apparatus. We report on the realization of a room-temperature $^{40}{\rm Ca}^{+}$ ion trapping vacuum apparatus with unprecedentedly low reaction rates of ions with a dominant vacuum contaminant: hydrogen. We present our trap assembly procedures and hydrogen pressure characterization by analysis of the CaH$^+$ molecule formation rate.
\end{abstract}

\maketitle

\section{Introduction}

Experimental platforms utilizing controllable electric potentials for storing several charged atomic particles have become pivotal in several modern research fields, including precision metrology, sensing, or quantum information processing~\cite{ludlow2015optical, monz2016realization, zhang2017observation}. Besides representing a pioneering platform for tests of a large number of advanced applications of quantum optics phenomena, they have also provided an indispensable resource for fundamental studies of atomic structure and its interaction with electromagnetic radiation. Dozens of experimental research groups worldwide utilize atomic ion traps as the main experimental platform and several efforts for the development of standardized technological substances with the possibility of involving automated industrial processing have emerged in past few years~\cite{salvador2019multilayer,niedermayr2014cryogenic}. One of the crucial technological challenges in process of realization of ion trapping apparatus corresponds to establishing sufficient vacuum environment in the trapping region. The achieved vacuum pressure is proportional to the mean free path of residual atomic particles, which in turn determines the achievable lifetime of the trapped ionic system. The collisions with the background gases can lead to thermalization of ions and their crystalized spatial structures, chemical reactions, or even direct ejection of ions from the trapping potential~\cite{micke2019closed,pagano2018cryogenic}. Vacuum quality can thus directly limit performed experiments and impose requirements on frequent ion reloading in some of the most advanced branches of ion trapping applications which involve large numbers of ions~\cite{zhang2017observation, friis2018observation, obvsil2018nonclassical, monz2014, keller2019controlling, kiethe2017probing, maier2019environment, ulm2013observation, bruzewicz2016scalable}.

The target vacuum pressures suitable for trapping single atomic ions or small ion crystals are in the ultra-high-vacuum (UHV) regime, well below $10^{-9}$\,mBar~\cite{pagano2018cryogenic, nizamani2013versatile}. Demands on the feasibility of scaling up the number of trapped ions further substantially decrease the required values of vacuum pressure levels, as the number of collisions scales linearly with the number of trapped ions. For thoroughly cleaned and well-baked chambers pre-pumped with a turbomolecular pump, the limiting residual element with highest partial pressure corresponds to a hydrogen due to its large concentration in stainless steel and high diffusion rate~\cite{calder1967reduction}. The chemical reactions of hydrogen with various atomic ion species were formerly studied due to their creation feasibility and suitability for studies of ultracold ion chemistry~\cite{carr2009cold,hansen2012single}, direct or sympathetic laser cooling~\cite{kimura2011sympathetic,chou2017preparation,rugango2016vibronic,staanum2010rotational,schneider2010all}, proposed as sensitive probes of fundamental theories~\cite{schiller2005tests}, or considered as astrophysically important molecules~\cite{petitprez1989infrared}.

Here we present our procedure of building an UHV apparatus containing a linear Paul trap for spatial localization of $^{40}$Ca$^+$ ions and estimation of the achieved hydrogen pressure using the measurement of the molecular reaction rates. The first part summarizes the particular employed procedures of vacuum chamber assembly, which are, up to small differences, well established in the experimental quantum optics community working with trapped cold atoms or ions. The resulting vacuum pressure level corresponds, to our best knowledge, to one of the best values reached in non-cryogenic vacuum apparatus in experimental trapped ion research and therefore we focus on the rigorous estimation of the pressure of the residual gas with the highest partial pressure corresponding to hydrogen molecules. The pressure is gauged by $^{40}$Ca$^+$ ion crystal by measurement and evaluation of reaction rates with the hydrogen background gas. The conclusion includes a comparison of presented results with recent works focused on the estimation of vacuum pressures from observable interaction rates of trapped ions with background gas.

\section{Vacuum construction}

The presented vacuum chamber is composed of and contains solely UHV compatible materials with low outgassing rates. The vacuum chamber and employed vacuum flanges are made from stainless steel SS~316~LN with low relative magnetic permeability in order to minimize unknown magnetic field gradients across the trapping region. For the same reason, all stainless steel components of the trap and its holder which mainly correspond to small screws are also made from SS~316~LN. The only exception corresponds to the combined vacuum pump (SAES group, NEXTorr$^\circledR$ D~100-5), which has SS~304 parts and includes a permanent magnet necessary for the operation of the ion pump, however, its position within the chamber corresponds to the largest distance to the trap. In addition to stainless steel, the large surface of the chamber is employed for optical access using a fused silica vacuum viewports with a broadband anti-reflection coatings covering 397 to 850~nm range. The trap and ion source assembly further contain Macor$^\circledR$, titanium, sapphire crystal, and an indium sealed source of calcium atoms. The trap electrodes are connected to oxygen-free high conductivity (OFHC) copper feedthroughs using an OFHC copper conductors and additional wires with Kapton$^\circledR$ isolation were used for the connection of the in-vacuum thermistor. Several parts were silver-plated, which included chamber assembly screws and majority of the OFHC copper gaskets. The total volume of the vacuum chamber has been estimated to approximately 1.6~litres.

The Paul trap itself is formed by six titanium electrodes, four radial and two axial, and additional two pairs of SS~316~LN stainless steel micro-motion compensation electrodes~\cite{guggemos2017}. All titanium electrodes were electroplated by approximately 10\,$\upmu$m thick gold layer to avoid any oxidized surfaces in the proximity of the trapping potential and, at the same time, enhance the electrical conductivity for the applied radio-frequency drive, which in turn results in lower temperature of the trap due to suppressed electric power dissipation. Before the electroplating, electrodes were cleaned in an ultrasonic bath with acetone and methanol, respectively. Subsequently, hydrogen fluoride was used to remove native titanium oxide. All etched parts were stored in the water in order to slow down any further oxidation. A gold sulfite electroplating solution TSG-250 heated to 60$^\circ$C was used for creation of satin-smooth electrodeposits on the titanium surface. The endcap electrodes were electroplated with 10~mA current for 40~minutes and radial electrodes with 5~mA for 30~minutes. The achieved gold surface has a low roughness with grain size of less than a micron in diameter. The trap design corresponds to ion to radial electrodes and ion to axial electrodes distances of 566~$\mu$m and 2250~$\mu$m, respectively. We note that the measured heating rates with single $^{40}{\rm Ca}^{+}$ ion using the thermometry on the 4S$_{\rm 1/2}\leftrightarrow3{\rm D}_{5/2}$ transition in the final assembled setup are $5.1\pm0.4$~phonon/s for the axial motion at~1.2~MHz.


Besides choosing the appropriate UHV compatible materials, the crucial point in the process of building any UHV vacuum chamber is cleanliness. Outgassing will play an important role below pressure levels of $10^{-6}$~mBar and potential residual organic compounds can limit the achievable vacuum pressure. During the whole cleaning and assembling process we were strictly using cleanroom suits, nitril powder free gloves certified for clean room applications, hair nets and anti-dust mouth masks. The setup was assembled in the regular optical laboratory inside a flow box employing the filter of class H13 of norm EN~1822. The parts of the vacuum chamber and pumping station assembly were cleaned in the ultrasonic (US) bath with industrial detergents, demineralized (DM) water and high-performance liquid chromatography (HPLC) grade alcohols, respectively. For the sake of comprehensiveness and repeatability, we specify the employed cleaning procedure in Table~\ref{tab:general_cleaning}.
\begin{table}[th]
\caption{\label{tab:general_cleaning}Overview of the general cleaning procedure used mainly for stainless steel parts.}
\begin{ruledtabular}
\begin{tabular}{ll}
1. & 10 min. in US bath in 5\% Simple Green$^\circledR$ \\
2. & Rinse under DM water to get rid of the detergent \\
3. & 5 min. in US bath in DM water \\
4. & Quick rinse (\textasciitilde 10 s) under DM water \\
5. & 10 min. in US bath in HPLC Aceton \\
6. & Quick rinse (\textasciitilde 10 s) under DM water \\
7. & 10 min. in US bath in HPLC Methanol\\
\end{tabular}
\end{ruledtabular}
\end{table}
The in-vacuum parts made from an OFHC copper tend to quickly oxidize on the surface when exposed to air and the oxide layers were removed by etching in diluted citric acid contained in industrial detergent Citranox$^\circledR$. The cleaned surface oxidizes fast in wet environment and it is important to avoid any contact with water or long exposure to air after final cleaning. Note that the employed cleaning procedure for copper parts presented in Tab.~\ref{tab:copper_cleaning} avoids the Aceton cleaning~\cite{kagwade2001photochemical}.
\begin{table}[h]
\caption{\label{tab:copper_cleaning}Overview of the copper cleaning procedure.}
\begin{ruledtabular}
\begin{tabular}{ll}
1. & 10-15 min. in US bath in 2\% Citranox$^\circledR$ \\
2. & Rapid (\textasciitilde 2 s) rinse under DM water to get rid of the detergent \\
&  and immediate submerge into HPLC Isopropyl alcohol \\
3. &10 min. in US bath in HPLC Isopropyl alcohol \\
4. & 10 min. in US bath in HPLC Methanol \\
\end{tabular}
\end{ruledtabular}
\end{table}
We didn't use any ultrasonic bath (US) cleaning for the employed vacuum chamber, electrical feedthroughs, and optical viewports, as they were specified to be clean upon delivery and so we just wiped them with HPLC methanol before the assembly. The gold-plated trap electrodes were merely submerged in the methanol bath. The thoroughly cleaned parts were always stored exclusively in flowbox during the periods of assembly procedures.

\begin{figure}[t!]
\begin{center}
\includegraphics[width=1.\columnwidth]{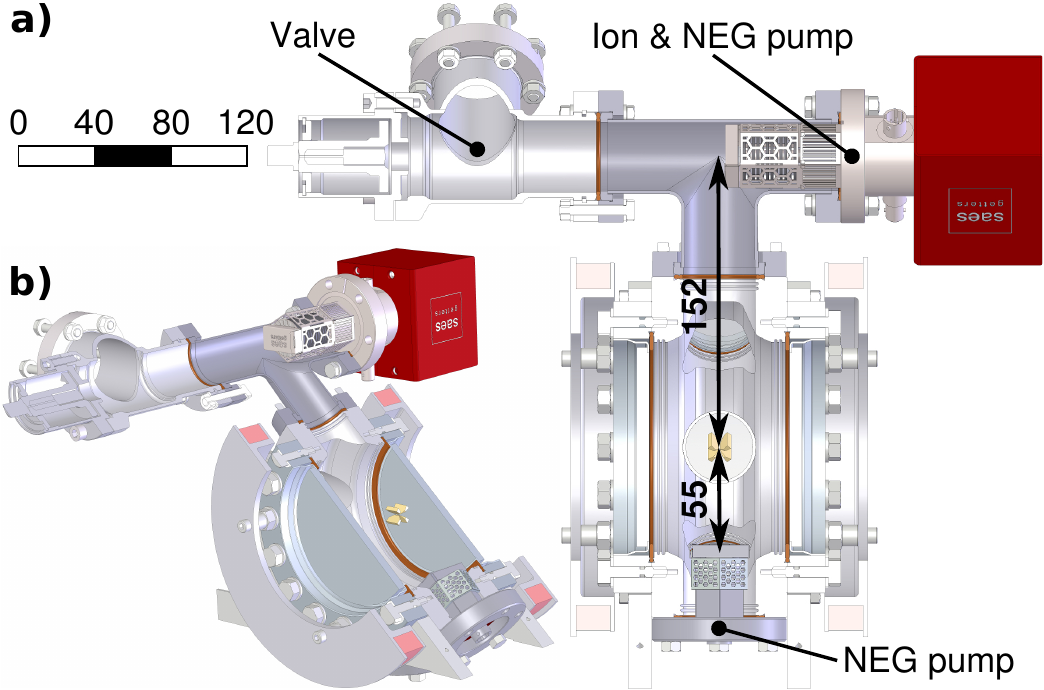}
\caption{A 3D-views of the employed vacuum chamber depict the relative positioning of the ion trap, entrapment pumps, and vacuum valve. a) The side-top projection with spatial scale in millimeters. b) The rotated perspective for illustration of the overall structure of the vacuum chamber.}
\label{fig:vacuum_design}
\end{center}
\end{figure}

Four different vacuum pumps were employed for the achievement of the resulting UHV pressures. Outside the vacuum chamber, a turbomolecular pump together with a Roots type pre-pump delivered as a single vacuum system (Pfeiffer Vacuum, HiCube 80 Pro/ACP 15) were used for reaching the UHV levels before activation of the entrapment pumps inside the chamber. The pumps in the vacuum chamber correspond to an ion pump combined with non-evaporative getter (NEG) pump (SAES Getters, NEXTorr$^\circledR$ D 100-5) and single NEG pump (SAES Getters, CapaciTorr$^\circledR$ D~50) positioned in the proximity of the ion trapping region at 55~mm distance, see Fig.~\ref{fig:vacuum_design}.

The vacuum level in a properly cleaned and outgassed apparatus is typically determined by hydrogen partial pressure, which diffuses through and from bulk stainless steel~\cite{calder1967reduction,chambers2004modern}. The conventional method of dealing with outgassing of hydrogen is based on establishment of a diffusion barrier on the surface of the stainless steel parts by an air-baking~\cite{westerberg1997hydrogen}. It has been predicted and observed that even small coating defects of the oxidized stainless steel flanges can lead to a lateral diffusion and outgassing of a hydrogen atoms into the chamber~\cite{prins1959}. The air-bake in a clean oven was realized after cleaning procedure with all stainless steel parts including the chamber, blank flanges, reduction and T-shape pipe~\cite{westerberg1997hydrogen}. The temperature was ramped up at a rate of 1$^\circ$C per minute to 400$^\circ$C, where it was maintained for 8~hours and afterwards ramped down at the same rate to reach room temperature. The air-bake resulted in \textbf{a} visible golden colour of the stainless steel surface corresponding to chromium-oxide.
A second baking step with simultaneous pumping of the preassembled chamber has been realized in order to outgas the hydrogen and water as much as possible before the final assembly. The chamber was assembled with blank stainless steel flanges instead of view-ports and the entrapment pumps and Paul trap were not installed, which allowed baking of the chamber at 270~$^\circ$C. After approximately one hour from the beginning of the pumping, a sufficiently low pressure had been established to perform a helium leak test which showed no detectable leaks. The ramp up and down speed of the temperature in the oven was set to the rate 15$^\circ$C per hour and the system was pumped at 270~$^\circ$C for 228~hours. During the whole baking time the partial gas pressures were monitored using the residual gas analyser (RGA) and a substantial decrease of the most prominent residual compounds, including predominant hydrogen, nitrogen, water, methanol and carbon dioxide, was observed. It is recommended to fill your vacuum system with pure nitrogen gas or at least dry air when opening the vacuum chamber, because outgassing of nitrogen is much easier than water adsorbed from humid air. The controlled filling of the chamber by high grade nitrogen also helps to avoid the risk of its contamination.
The third baking included the final assembly of ion trapping apparatus and its temperature was limited by optical viewports to 200$^\circ$C. Similarly as in the second baking stage, the helium leak test has been performed soon after starting the pumping procedure and then the temperature inside the baking oven was ramped up at a rate of 15~$^\circ$C per hour to the target temperature of 195~$^\circ$C where it stayed for 10~days. The ramp down at the same rate has been targeted to 45~$^\circ$C at which the ion pump magnets were attached and the pump was running for 1~minute for degassing. After removal of the magnets, the atomic oven containing calcium metal powder originally closed under an argon atmosphere and sealed with indium plug was heated for 2~minutes to degas. Last procedure realized inside the baking oven was the activation of NEG pumps using the original commercial driving units. One of the NEGs was first started in the conditioning regime to avoid a rapid increase of pressure in the chamber and after about 2~minutes also the conditioning of the second NEG pump was switched on. The actual pressure was observed on RGA and after 15~minutes the full activation power was switched on for the first NEG pump and with a 2~minutes delay also the second. Both NEG pumps were activated for 1~hour and after that, both driving units and baking oven were switched off. After thermalization of the NEG pumps, the ion pump was switched on and the valve between chamber and turbomolecular pump was closed a few minutes afterwards, while the chamber temperature being still at about 40~$^\circ$C.

The last partial pressures before the valve closure estimated at RGA were signifying the dominant residual gas being hydrogen with the partial pressure of $(4.62\pm0.05)\,\cdot 10^{-8}$~mBar followed by more than an order of magnitude lower population of nitrogen/carbon monoxide $(1.5\pm0.2)\,\cdot 10^{-9}$~mBar, carbon dioxide $(1.3\pm0.2)\,\cdot 10^{-9}$~mBar, water $(1.0\pm0.2)\,\cdot\,10^{-9}$~mBar and methanol $(6\pm2)\,\cdot\,10^{-10}$~mBar, with the rest of the residuals being below the noise floor of the RGA at about $2\cdot\,10^{-10}$~mBar level. The valve closure has resulted in a step increase of hydrogen pressure on RGA outside the chamber, which was the clear sign of its pumping by NEGs. The current on the ion pump monitor decreased quickly and after approximately two hours it showed solely 0~nA corresponding to the pressure level below its gauging capabilities, which is specified to be limited to about $<10^{-11}$~mBar for nitrogen gas.

\section{Estimation of the hydrogen pressure}

The attainable information about the pressure level inside the chamber using the ion pump gauge is rough and gives limited knowledge about the quality of achieved vacuum. Unfortunately, it corresponds to the only available instrument for vacuum pressure estimation in our setup, similarly as for most of apparatus in the ion trapping community.
The basic technical prerequisite for operation of many trapped ion experiments is the sufficiently long lifetime of the laser-cooled crystallized ion ensemble, which is typically limited by the surrounding vacuum level, laser cooling efficiency, or physical properties of the particular employed trap. The dependence of the ion crystal lifetime on the vacuum level can be in turn used for the precise estimation of the vacuum level and, together with the knowledge of particular reaction rate constants of given atomic isotope with prominent vacuum contaminants, serve also for the residual gas analysis. The conventional methods for vacuum pressure analysis using trapped atomic ions are based on estimation of rates of two types of interaction with background gas. One is corresponding to inelastic collisions leading to the formation of molecular ions~\cite{pagano2018cryogenic}, and the other is involving the elastic momentum transfer, which can result in an exchange of ion positions, rapid thermalization of ion crystal, or even loss of ions from the trap~\cite{hankin2019systematic}. The relatively high depth of our trapping potential estimated at 3.5~eV prevents a direct escape of ions and we do not observe any ion loss in usual experimental settings. At the same time, unambiguous measurement of ion crystal thermalization rates can be tricky due to short Doppler re-cooling times. For these reasons, elastic collision rates with background gas are typically measured by observation of crystal structure changes with one or more non-fluorescing ions purposely loaded in the string~\cite{hankin2019systematic,hempel2014digital}. However, such method is highly dependent on particular trapping parameters which determine the potential barrier between observed crystal orders, electric field noise properties, or motional state of ions and probability of thermal activation of position flips~\cite{hankin2019systematic,pagano2018cryogenic}. On the other hand, the vacuum analysis based on the observation of molecular reaction rates is highly insensitive to particular trapping and laser cooling settings and can thus lead to smaller systematic uncertainties of the estimated pressures, provided that the type of the molecular association reaction and corresponding rate coefficients are well known.

\begin{figure}[t!]
\begin{center}
\includegraphics[width=1.\columnwidth]{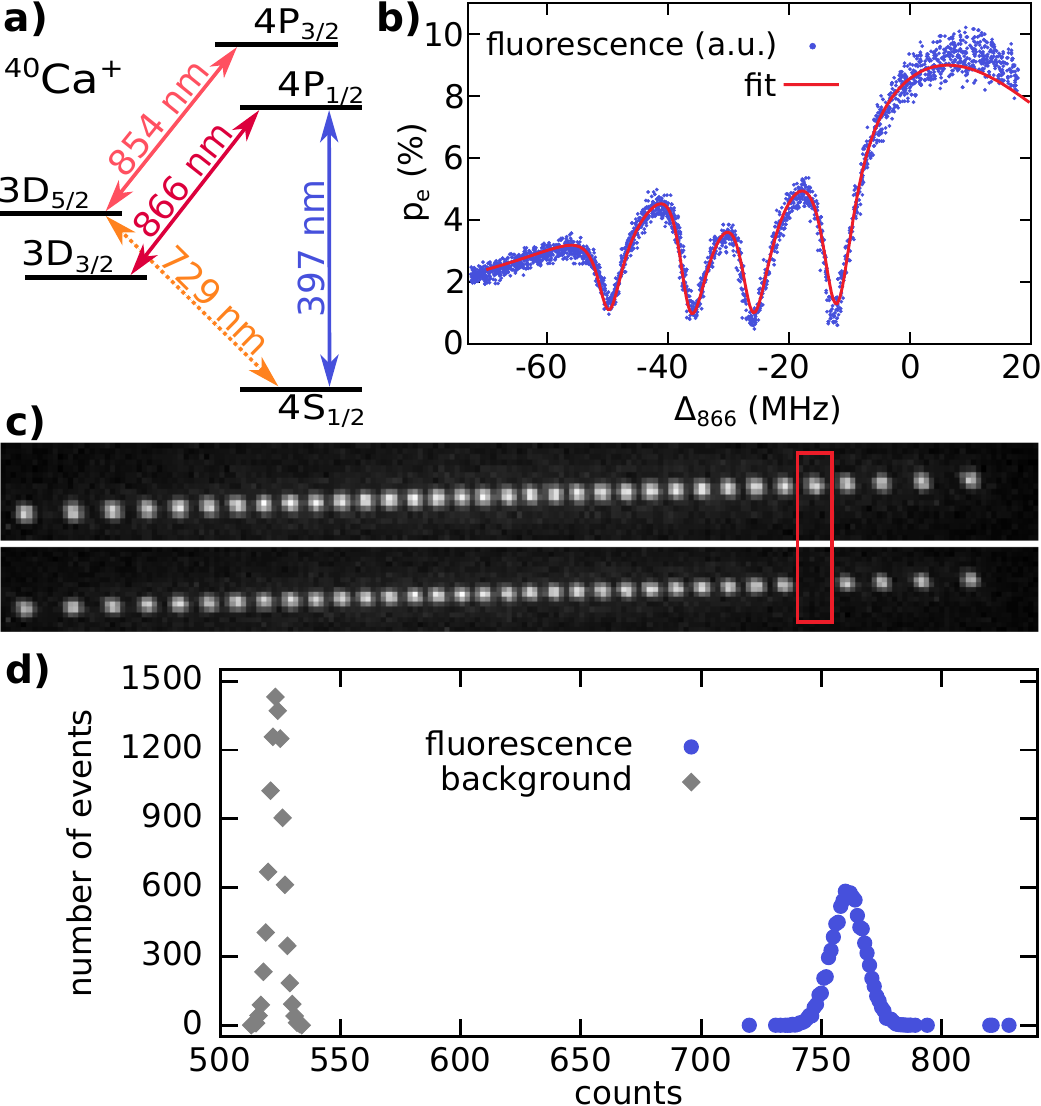}
\caption{a) A simplified energy level scheme of $^{40}$Ca$^+$ with corresponding excitation laser wavelengths. b) Measured and fitted fluorescence spectra as a function of the repumping 866~nm laser detuning for calibration of the 4P$_{1/2}$ state population and corresponding detected EMCCD camera intensity. c) The representative images from EMCCD camera showing full 34~ion string (top), and the rare event of string with a single ion becoming non-fluorescing due to the chemical reaction (bottom). d) The evaluated number of detection events with measured intensities of detected fluorescence for the outermost ion in the 34-ion string. The histogram is evaluated from $10^4$~measurements on the EMCCD camera, each with an acquisition time of 0.5~s. The large difference between the detected bright ion intensity distribution (blue circles) and background intensity values (grey squares) ensures the unambiguous detection of dark ion production even at timescales shorter than half of our acquisition period.}
\label{fig:vacuum_est}
\end{center}
\end{figure}

The presented vacuum chamber contains  $^{40}$Ca$^+$ ion trapping apparatus and the hydrogen partial pressure can thus be estimated by measurement of the rate of the reaction~\cite{kimura2011sympathetic, hansen2012single}
\begin{equation}
^{40}{\rm Ca}^+(4{\rm P}_{1/2}) + {\rm H}_2 \rightarrow ^{40}{\rm CaH}^+ + {\rm H},
\label{eq:chemistry}
\end{equation}
with its dynamics given by
\begin{equation}
N(^{40}\text{Ca}^+)=N_0(^{40}\text{Ca}^+) \cdot e^{-\gamma \cdot t}.
\label{eq:rate}
\end{equation}
Here, $\gamma$ is the reaction rate, $N_0(^{40}\text{Ca}^+)$ is the number of non-reacted ions at the beginning of the measurement period $t=0$, and $N(^{40}\text{Ca}^+)$ is the number of non-reacted ions after time $t$. The reaction rate $\gamma=p_{e}n_{{\rm H}_2}k_r$, with $p_{e}$ being the probability of 4P$_{1/2}$ level population, $n_{{\rm H}_2}$ the number density of the ${\rm H}_2$ gas, and $k_r$ is the reaction rate constant.

We trap linear $^{40}$Ca$^+$ ion strings in the installed linear Paul trap operated with 30.2~MHz radio-frequency drive and 3~W power resulting in radial center of mass motional frequencies of $f_x \approx f_y \approx 1.86$~MHz. The axial potential is set for optimizing the stability of the employed linear ion string, which for the hydrogen pressure analysing measurements with 34~ions corresponds to the center of mass mode frequency $f_z \approx 91$~kHz and $U_{\rm tip}=8$~V. The micromotion has been carefully optimized for employed trapping potential settings. The Doppler cooling of ions is realized by a 397~nm laser red detuned from the 4S$_{1/2}\leftrightarrow$ 4P$_{1/2}$ transition and additional 866~nm laser beam is used for depopulation of the metastable 3D$_{3/2}$ manifold, see Fig.~\ref{fig:vacuum_est}-a). The 854~nm laser beam set to the 3D$_{5/2}\leftrightarrow$ 4P$_{3/2}$ transition resonance is used for depopulation of the 3D$_{5/2}$ level, which is populated by extremely rare events corresponding to 397~nm laser excitation of the 4S$_{1/2}\leftrightarrow$ 4P$_{3/2}\leftrightarrow$ 3D$_{5/2}$ Raman transition.
The 397~nm fluorescence scattered from ions in the direction perpendicular to the symmetry axis of the employed linear Paul trap is collected by a lens positioned outside the vacuum chamber and observed on EMCCD camera. The measured fluorescence emission spectrum with four dark resonances shown in the Fig.~\ref{fig:vacuum_est}-b) serves for precise calibration of the 4P$_{1/2}$ population $p_{e}$ using the steady state solution of 8-level optical Bloch equations. The standard Doppler cooling conditions correspond to $p_{e}$ on the order of a few percent and the chemical reaction rate is too small to be observable in our vacuum level even with several tens of ions. In order to homogeneously enhance the laser induced reaction rate~(\ref{eq:chemistry}), the probability $p_{e}$ was further tuned up to the average value of $p_{e}=(15.9 \pm 1.7)\%$ for all ions in the crystal using an additional 397~nm excitation beam propagating along the ion string. Nevertheless, the observation of a few chemical reactions still necessitated very long measurements and thus stable laser excitation parameters were ensured by frequency and intensity stabilizations of both lasers, with frequency locks employing as a reference single optical frequency comb referenced to hydrogen maser.

The illustration of the ion crystal employed in the presented measurement containing 34~$^{40}\text{Ca}^+$ ions is shown in the Fig.~\ref{fig:vacuum_est}-c), where the bottom picture illustrates one of the extremely rare events of one ion getting dark. This crystal has been observed for~51 hours and its image has been captured with 1.95~Hz rate and 500~ms acquisition time.
The post-processing of the recorded images showed in total 4~reactions.
After becoming dark, the ion always stayed in non-fluorescing state on a timescale of minutes, namely for 63.5~s, 934.5~s,  506~s and 366.5~s and then became bright again, which can be most likely attributed to an off-resonant optical dissociation~\cite{rugango2016vibronic}. The time-lapse images were processed one by one by averaging the photon counts from pixels surrounding the ion image for each ion separately. A typical evaluated histogram of detected count rates shown in Fig.~\ref{fig:vacuum_est}-d) demonstrates that the fluorescence peak for each ion is separated very well from the background count rate values and thus any nonfluorescing events longer than 0.25~s would be unambiguously recognized.
The observed timescales of the three dissociation events statistically practically exclude any dark events where the dissociation time would be faster or comparable to 0.25~s in the same experimental conditions, which further justifies the evaluation of the reaction rate as 1~reaction per 12.75~hours. The corresponding partial pressure of the ${\rm H}_2$ in our system can be found using the equation~(\ref{eq:rate}), which results in $P_{\rm H_2} \leq (2.1\pm 1.1)\cdot 10^{-13}$~mBar and $P_{\rm H_2} \leq (5.5\pm 2.8)\cdot 10^{-12}$~mBar, when employing the reaction rate coefficient $k_r$ from~\cite{kimura2011sympathetic} and from~\cite{hansen2012single}, respectively. The quoted standard deviations correspond to two statistically independent uncertainty contributions. The dominant contribution is caused by the uncertainty of random and time independent success of observation of molecular reaction. The variance of number of reaction events is given by the Poisson distribution with a very large number of trials and very small success probability corresponding to 4 events per total number of trials. This value has been convolved with statistically independent uncertainty of 4P$_{1/2}$ state population, which has been found by the statistical analysis of the detected fluorescence count rate.

The provided partial pressure analysis assumes that all molecular association events correspond to reaction of $^{40}$Ca$^+$ with ${\rm H}_2$. This assumption is based on the dominant residual content of hydrogen observable on RGA before the chamber closure and on the typical characteristics of the residual gas content in the vast majority of stainless steel UHV chambers~\cite{hankin2019systematic,pagano2018cryogenic,micke2019closed,chambers2004modern,calder1967reduction}. Although it is unlikely that we see reactions with other vacuum contaminants, it is practically demanding to implement means to unambiguously prove it and we have to regard the presented values as an upper limit on the hydrogen partial pressure, as contribution of any other element to the observed dark ion creation would effectively correspond to smaller hydrogen content in the chamber.

\section{Discussion}

The ultra-high-vacuum generation corresponds to a routine, yet the crucial technological task necessary for achieving control over single charged atomic ions and its methods are continuously revisited and improved among whole ion trapping community. Current efforts on scaling some prominent trapped-ion applications require experimenting with stable large ionic arrays which tighten up already stringent requirements on the vacuum levels well below~$10^{-9}$~mBar. A room temperature ion trapping chamber presented here has recently demonstrated the distinctive ability of working with large and stable ion ensembles in experiments, where temporal fluctuations of ion numbers would inevitably cause a severe reduction of the observed phenomena~\cite{obvsil2018nonclassical, obvsil2018scalable}. The rigorous estimation of the achieved hydrogen partial pressure using the estimation of molecular reaction rates has led to its upper limits $P_{\rm H_2} \leq (5.5\pm 2.8)\cdot 10^{-12}$~mBar and $P_{\rm H_2} \leq (2.1\pm 1.1)\cdot 10^{-13}$~mBar using the reaction rate coefficient from~\cite{hansen2012single} and from~\cite{kimura2011sympathetic}. This corresponds to four chemical reactions in 51~hours long measurement with 34~ion crystal, which, to the best of our knowledge, is the lowest value reported for a room-temperature ion trapping experiment. The dominant uncertainty in the observed upper limit on the hydrogen partial pressure is caused by the large discrepancy in the reported reaction rate constants, which could not be feasibly resolved in the current vacuum apparatus. We plan to implement measures in a new vacuum chamber with Ca$^+$ ion trap comprising a high precision residual gas analyser, which would allow for its comparison.

The provided explicit instructions for the vacuum generation, hydrogen pressure estimation, and the achieved vacuum level itself, do not intend to be recognized as a milestone breaking, but rather constitute a solid and reliable reference for the realization and calibration of a chamber capable of temporarily stable storage of large ionic arrays.
They complement a rich variety of recent notable reports from ion trapping community on measurement of effective lifetimes of ion strings with atomic ions localized in Paul traps. We do not attempt to quantitatively compare the reported vacuum levels, as different measures and atomic species could in principle lead to incomplete conclusions. We instead enlist the recent reported measurements relevant for the potential applications. The collision rates on the order of a single collision per hour per trapped ion have been reported by observations of dark ion position change in the $^{171}$Yb$^+$ ion string~\cite{zhang2017observation}. The same setup reported an average lifetime of about 5~minutes for 53~ion crystal with rare events reaching up to 30~minutes. Measurements of rates of the dark ion position change in the $^{40}$Ca$^+$ crystal resulted in single ion position change every 86~s on average for a 6-ion crystal and every 27~s for a chain of 20~ions~\cite{hempel2014digital}. The same method employed for a 5-ion crystal of $^{171}$Yb$^+$ resulted in hopping every 49~s on average~\cite{furst2019trapped}. The recent rigorous analysis of reorder rate for a two-ion $^{40}$Ca$^+ - ^{27}$Al$^+$ crystal motivated by the estimation of frequency shift in the optical atomic clock caused by background gas collisions resulted in the hydrogen-dominated background gas pressure of $(5.47\pm 0.08)\times 10^{-10}$~mBar~\cite{hankin2019systematic}.  We stress again, that these results are generally dependent on trapping and laser cooling parameters, and thus not directly comparable.
Alternative approach for achieving very low vacuum levels employs cryogenic vacuum setups~\cite{pagano2018cryogenic,micke2019closed}. Besides simplifying and speeding up the vacuum construction due to suppressed desorption rates, the reduced kinetic energy of the background gases limits the observable heating of the ion crystal caused by collisions. The background gas pressure of P$_{4{\rm K}}<1.3\cdot 10^{-13}$~mBar was estimated from dark ion production rates, while counting of the sensitive zig-zag crystal structural changes gave~P$_{4.7{\rm K}}<1.3\cdot 10^{-12}$~mBar~\cite{pagano2018cryogenic}. Cryogenic setup reported in reference~\cite{micke2019closed} employed highly charged ions and the charge-exchange reactions with hydrogen to derive its upper vacuum pressure limit of~P$_{4.6{\rm K}}<1.26 (-0.11/+0.12)\cdot 10^{-14}$~mBar. We believe that, together with these recent remarkable achievements, presented guidelines for the ion trapping vacuum chamber construction and achieved pressure estimation will facilitate current efforts in mastering large trapped ion arrays for a broad range of applications of quantum technologies~\cite{zhang2017observation, friis2018observation, obvsil2018nonclassical, monz2014, keller2019controlling, kiethe2017probing, maier2019environment, ulm2013observation}.

\begin{acknowledgments}
We acknowledge the kind technological support from the group of Rainer Blatt including the contribution of Stefan Haslwanter to the production of the employed Paul trap electrodes and Yves Colombe for valuable advice about the vacuum design and construction. We thank Darren Moore for careful reading of the manuscript.
This work has been supported by the grant No. GA19-14988S of the Czech Science Foundation, CZ.02.1.01/0.0/0.0/16\_026/0008460 of MEYS CR and Palacky University IGA-PrF-2019-010. The research used infrastructure by MEYS CR, EC, and CAS (LO1212, CZ.1.05/2.1.00/01.0017, RVO:68081731).
\end{acknowledgments}


\begin{thebibliography}{35}%
\makeatletter
\providecommand \@ifxundefined [1]{%
 \@ifx{#1\undefined}
}%
\providecommand \@ifnum [1]{%
 \ifnum #1\expandafter \@firstoftwo
 \else \expandafter \@secondoftwo
 \fi
}%
\providecommand \@ifx [1]{%
 \ifx #1\expandafter \@firstoftwo
 \else \expandafter \@secondoftwo
 \fi
}%
\providecommand \natexlab [1]{#1}%
\providecommand \enquote  [1]{``#1''}%
\providecommand \bibnamefont  [1]{#1}%
\providecommand \bibfnamefont [1]{#1}%
\providecommand \citenamefont [1]{#1}%
\providecommand \href@noop [0]{\@secondoftwo}%
\providecommand \href [0]{\begingroup \@sanitize@url \@href}%
\providecommand \@href[1]{\@@startlink{#1}\@@href}%
\providecommand \@@href[1]{\endgroup#1\@@endlink}%
\providecommand \@sanitize@url [0]{\catcode `\\12\catcode `\$12\catcode
  `\&12\catcode `\#12\catcode `\^12\catcode `\_12\catcode `\%12\relax}%
\providecommand \@@startlink[1]{}%
\providecommand \@@endlink[0]{}%
\providecommand \url  [0]{\begingroup\@sanitize@url \@url }%
\providecommand \@url [1]{\endgroup\@href {#1}{\urlprefix }}%
\providecommand \urlprefix  [0]{URL }%
\providecommand \Eprint [0]{\href }%
\providecommand \doibase [0]{http://dx.doi.org/}%
\providecommand \selectlanguage [0]{\@gobble}%
\providecommand \bibinfo  [0]{\@secondoftwo}%
\providecommand \bibfield  [0]{\@secondoftwo}%
\providecommand \translation [1]{[#1]}%
\providecommand \BibitemOpen [0]{}%
\providecommand \bibitemStop [0]{}%
\providecommand \bibitemNoStop [0]{.\EOS\space}%
\providecommand \EOS [0]{\spacefactor3000\relax}%
\providecommand \BibitemShut  [1]{\csname bibitem#1\endcsname}%
\let\auto@bib@innerbib\@empty
\bibitem [{\citenamefont {Ludlow}\ \emph {et~al.}(2015)\citenamefont {Ludlow},
  \citenamefont {Boyd}, \citenamefont {Ye}, \citenamefont {Peik},\ and\
  \citenamefont {Schmidt}}]{ludlow2015optical}%
  \BibitemOpen
  \bibfield  {author} {\bibinfo {author} {\bibfnamefont {A.~D.}\ \bibnamefont
  {Ludlow}}, \bibinfo {author} {\bibfnamefont {M.~M.}\ \bibnamefont {Boyd}},
  \bibinfo {author} {\bibfnamefont {J.}~\bibnamefont {Ye}}, \bibinfo {author}
  {\bibfnamefont {E.}~\bibnamefont {Peik}}, \ and\ \bibinfo {author}
  {\bibfnamefont {P.~O.}\ \bibnamefont {Schmidt}},\ }\bibfield  {title}
  {\enquote {\bibinfo {title} {Optical atomic clocks},}\ }\href
  {https://journals.aps.org/rmp/abstract/10.1103/RevModPhys.87.637} {\bibfield
  {journal} {\bibinfo  {journal} {Reviews of Modern Physics}\ }\textbf
  {\bibinfo {volume} {87}},\ \bibinfo {pages} {637} (\bibinfo {year}
  {2015})}\BibitemShut {NoStop}%
\bibitem [{\citenamefont {Monz}\ \emph {et~al.}(2016)\citenamefont {Monz},
  \citenamefont {Nigg}, \citenamefont {Martinez}, \citenamefont {Brandl},
  \citenamefont {Schindler}, \citenamefont {Rines}, \citenamefont {Wang},
  \citenamefont {Chuang},\ and\ \citenamefont {Blatt}}]{monz2016realization}%
  \BibitemOpen
  \bibfield  {author} {\bibinfo {author} {\bibfnamefont {T.}~\bibnamefont
  {Monz}}, \bibinfo {author} {\bibfnamefont {D.}~\bibnamefont {Nigg}}, \bibinfo
  {author} {\bibfnamefont {E.~A.}\ \bibnamefont {Martinez}}, \bibinfo {author}
  {\bibfnamefont {M.~F.}\ \bibnamefont {Brandl}}, \bibinfo {author}
  {\bibfnamefont {P.}~\bibnamefont {Schindler}}, \bibinfo {author}
  {\bibfnamefont {R.}~\bibnamefont {Rines}}, \bibinfo {author} {\bibfnamefont
  {S.~X.}\ \bibnamefont {Wang}}, \bibinfo {author} {\bibfnamefont {I.~L.}\
  \bibnamefont {Chuang}}, \ and\ \bibinfo {author} {\bibfnamefont
  {R.}~\bibnamefont {Blatt}},\ }\bibfield  {title} {\enquote {\bibinfo {title}
  {Realization of a scalable shor algorithm},}\ }\href
  {http://science.sciencemag.org/content/351/6277/1068.short} {\bibfield
  {journal} {\bibinfo  {journal} {Science}\ }\textbf {\bibinfo {volume}
  {351}},\ \bibinfo {pages} {1068--1070} (\bibinfo {year} {2016})}\BibitemShut
  {NoStop}%
\bibitem [{\citenamefont {Zhang}\ \emph {et~al.}(2017)\citenamefont {Zhang},
  \citenamefont {Pagano}, \citenamefont {Hess}, \citenamefont {Kyprianidis},
  \citenamefont {Becker}, \citenamefont {Kaplan}, \citenamefont {Gorshkov},
  \citenamefont {Gong},\ and\ \citenamefont {Monroe}}]{zhang2017observation}%
  \BibitemOpen
  \bibfield  {author} {\bibinfo {author} {\bibfnamefont {J.}~\bibnamefont
  {Zhang}}, \bibinfo {author} {\bibfnamefont {G.}~\bibnamefont {Pagano}},
  \bibinfo {author} {\bibfnamefont {P.~W.}\ \bibnamefont {Hess}}, \bibinfo
  {author} {\bibfnamefont {A.}~\bibnamefont {Kyprianidis}}, \bibinfo {author}
  {\bibfnamefont {P.}~\bibnamefont {Becker}}, \bibinfo {author} {\bibfnamefont
  {H.}~\bibnamefont {Kaplan}}, \bibinfo {author} {\bibfnamefont {A.~V.}\
  \bibnamefont {Gorshkov}}, \bibinfo {author} {\bibfnamefont {Z.-X.}\
  \bibnamefont {Gong}}, \ and\ \bibinfo {author} {\bibfnamefont
  {C.}~\bibnamefont {Monroe}},\ }\bibfield  {title} {\enquote {\bibinfo {title}
  {Observation of a many-body dynamical phase transition with a 53-qubit
  quantum simulator},}\ }\href {https://www.nature.com/articles/nature24654}
  {\bibfield  {journal} {\bibinfo  {journal} {Nature}\ }\textbf {\bibinfo
  {volume} {551}},\ \bibinfo {pages} {601} (\bibinfo {year}
  {2017})}\BibitemShut {NoStop}%
\bibitem [{\citenamefont {Salvador}\ \emph {et~al.}(2019)\citenamefont
  {Salvador}, \citenamefont {Zarantonello}, \citenamefont {Hahn}, \citenamefont
  {Preciado-Grijalva}, \citenamefont {Morgner}, \citenamefont {Wahnschaffe},\
  and\ \citenamefont {Ospelkaus}}]{salvador2019multilayer}%
  \BibitemOpen
  \bibfield  {author} {\bibinfo {author} {\bibfnamefont {A.~B.}\ \bibnamefont
  {Salvador}}, \bibinfo {author} {\bibfnamefont {G.}~\bibnamefont
  {Zarantonello}}, \bibinfo {author} {\bibfnamefont {H.}~\bibnamefont {Hahn}},
  \bibinfo {author} {\bibfnamefont {A.}~\bibnamefont {Preciado-Grijalva}},
  \bibinfo {author} {\bibfnamefont {J.}~\bibnamefont {Morgner}}, \bibinfo
  {author} {\bibfnamefont {M.}~\bibnamefont {Wahnschaffe}}, \ and\ \bibinfo
  {author} {\bibfnamefont {C.}~\bibnamefont {Ospelkaus}},\ }\bibfield  {title}
  {\enquote {\bibinfo {title} {Multilayer ion trap technology for scalable
  quantum computing and quantum simulation},}\ }\href
  {https://iopscience.iop.org/article/10.1088/1367-2630/ab0e46/meta} {\bibfield
   {journal} {\bibinfo  {journal} {New Journal of Physics}\ }\textbf {\bibinfo
  {volume} {21}},\ \bibinfo {pages} {043011} (\bibinfo {year}
  {2019})}\BibitemShut {NoStop}%
\bibitem [{\citenamefont {Niedermayr}\ \emph {et~al.}(2014)\citenamefont
  {Niedermayr}, \citenamefont {Lakhmanskiy}, \citenamefont {Kumph},
  \citenamefont {Partel}, \citenamefont {Edlinger}, \citenamefont {Brownnutt},\
  and\ \citenamefont {Blatt}}]{niedermayr2014cryogenic}%
  \BibitemOpen
  \bibfield  {author} {\bibinfo {author} {\bibfnamefont {M.}~\bibnamefont
  {Niedermayr}}, \bibinfo {author} {\bibfnamefont {K.}~\bibnamefont
  {Lakhmanskiy}}, \bibinfo {author} {\bibfnamefont {M.}~\bibnamefont {Kumph}},
  \bibinfo {author} {\bibfnamefont {S.}~\bibnamefont {Partel}}, \bibinfo
  {author} {\bibfnamefont {J.}~\bibnamefont {Edlinger}}, \bibinfo {author}
  {\bibfnamefont {M.}~\bibnamefont {Brownnutt}}, \ and\ \bibinfo {author}
  {\bibfnamefont {R.}~\bibnamefont {Blatt}},\ }\bibfield  {title} {\enquote
  {\bibinfo {title} {Cryogenic surface ion trap based on intrinsic silicon},}\
  }\href
  {https://iopscience.iop.org/article/10.1088/1367-2630/16/11/113068/meta}
  {\bibfield  {journal} {\bibinfo  {journal} {New Journal of Physics}\ }\textbf
  {\bibinfo {volume} {16}},\ \bibinfo {pages} {113068} (\bibinfo {year}
  {2014})}\BibitemShut {NoStop}%
\bibitem [{\citenamefont {Micke}\ \emph {et~al.}(2019)\citenamefont {Micke},
  \citenamefont {Stark}, \citenamefont {King}, \citenamefont {Leopold},
  \citenamefont {Pfeifer}, \citenamefont {Schm{\"o}ger}, \citenamefont
  {Schwarz}, \citenamefont {Spie{\ss}}, \citenamefont {Schmidt},\ and\
  \citenamefont {L{\'o}pez-Urrutia}}]{micke2019closed}%
  \BibitemOpen
  \bibfield  {author} {\bibinfo {author} {\bibfnamefont {P.}~\bibnamefont
  {Micke}}, \bibinfo {author} {\bibfnamefont {J.}~\bibnamefont {Stark}},
  \bibinfo {author} {\bibfnamefont {S.~A.}\ \bibnamefont {King}}, \bibinfo
  {author} {\bibfnamefont {T.}~\bibnamefont {Leopold}}, \bibinfo {author}
  {\bibfnamefont {T.}~\bibnamefont {Pfeifer}}, \bibinfo {author} {\bibfnamefont
  {L.}~\bibnamefont {Schm{\"o}ger}}, \bibinfo {author} {\bibfnamefont
  {M.}~\bibnamefont {Schwarz}}, \bibinfo {author} {\bibfnamefont
  {L.}~\bibnamefont {Spie{\ss}}}, \bibinfo {author} {\bibfnamefont {P.~O.}\
  \bibnamefont {Schmidt}}, \ and\ \bibinfo {author} {\bibfnamefont
  {J.}~\bibnamefont {L{\'o}pez-Urrutia}},\ }\bibfield  {title} {\enquote
  {\bibinfo {title} {Closed-cycle, low-vibration 4 k cryostat for ion traps and
  other applications},}\ }\href {https://arxiv.org/abs/1901.03630} {\bibfield
  {journal} {\bibinfo  {journal} {arXiv preprint arXiv:1901.03630}\ } (\bibinfo
  {year} {2019})}\BibitemShut {NoStop}%
\bibitem [{\citenamefont {Pagano}\ \emph {et~al.}(2018)\citenamefont {Pagano},
  \citenamefont {Hess}, \citenamefont {Kaplan}, \citenamefont {Tan},
  \citenamefont {Richerme}, \citenamefont {Becker}, \citenamefont
  {Kyprianidis}, \citenamefont {Zhang}, \citenamefont {Birckelbaw},
  \citenamefont {Hernandez}, \citenamefont {Wu},\ and\ \citenamefont
  {Monroe}}]{pagano2018cryogenic}%
  \BibitemOpen
  \bibfield  {author} {\bibinfo {author} {\bibfnamefont {G.}~\bibnamefont
  {Pagano}}, \bibinfo {author} {\bibfnamefont {P.~W.}\ \bibnamefont {Hess}},
  \bibinfo {author} {\bibfnamefont {H.~B.}\ \bibnamefont {Kaplan}}, \bibinfo
  {author} {\bibfnamefont {W.~L.}\ \bibnamefont {Tan}}, \bibinfo {author}
  {\bibfnamefont {P.}~\bibnamefont {Richerme}}, \bibinfo {author}
  {\bibfnamefont {P.}~\bibnamefont {Becker}}, \bibinfo {author} {\bibfnamefont
  {A.}~\bibnamefont {Kyprianidis}}, \bibinfo {author} {\bibfnamefont
  {J.}~\bibnamefont {Zhang}}, \bibinfo {author} {\bibfnamefont
  {E.}~\bibnamefont {Birckelbaw}}, \bibinfo {author} {\bibfnamefont {M.~R.}\
  \bibnamefont {Hernandez}}, \bibinfo {author} {\bibfnamefont {Y.}~\bibnamefont
  {Wu}}, \ and\ \bibinfo {author} {\bibfnamefont {C.}~\bibnamefont {Monroe}},\
  }\bibfield  {title} {\enquote {\bibinfo {title} {Cryogenic trapped-ion system
  for large scale quantum simulation},}\ }\href {\doibase
  10.1088/2058-9565/aae0fe} {\bibfield  {journal} {\bibinfo  {journal} {Quantum
  Science and Technology}\ }\textbf {\bibinfo {volume} {4}},\ \bibinfo {pages}
  {014004} (\bibinfo {year} {2018})}\BibitemShut {NoStop}%
\bibitem [{\citenamefont {Friis}\ \emph {et~al.}(2018)\citenamefont {Friis},
  \citenamefont {Marty}, \citenamefont {Maier}, \citenamefont {Hempel},
  \citenamefont {Holz{\"a}pfel}, \citenamefont {Jurcevic}, \citenamefont
  {Plenio}, \citenamefont {Huber}, \citenamefont {Roos}, \citenamefont
  {Blatt},\ and\ \citenamefont {Lanyon}}]{friis2018observation}%
  \BibitemOpen
  \bibfield  {author} {\bibinfo {author} {\bibfnamefont {N.}~\bibnamefont
  {Friis}}, \bibinfo {author} {\bibfnamefont {O.}~\bibnamefont {Marty}},
  \bibinfo {author} {\bibfnamefont {C.}~\bibnamefont {Maier}}, \bibinfo
  {author} {\bibfnamefont {C.}~\bibnamefont {Hempel}}, \bibinfo {author}
  {\bibfnamefont {M.}~\bibnamefont {Holz{\"a}pfel}}, \bibinfo {author}
  {\bibfnamefont {P.}~\bibnamefont {Jurcevic}}, \bibinfo {author}
  {\bibfnamefont {M.~B.}\ \bibnamefont {Plenio}}, \bibinfo {author}
  {\bibfnamefont {M.}~\bibnamefont {Huber}}, \bibinfo {author} {\bibfnamefont
  {C.}~\bibnamefont {Roos}}, \bibinfo {author} {\bibfnamefont {R.}~\bibnamefont
  {Blatt}}, \ and\ \bibinfo {author} {\bibfnamefont {B.}~\bibnamefont
  {Lanyon}},\ }\bibfield  {title} {\enquote {\bibinfo {title} {Observation of
  entangled states of a fully controlled 20-qubit system},}\ }\href
  {https://journals.aps.org/prx/abstract/10.1103/PhysRevX.8.021012} {\bibfield
  {journal} {\bibinfo  {journal} {Physical Review X}\ }\textbf {\bibinfo
  {volume} {8}},\ \bibinfo {pages} {021012} (\bibinfo {year}
  {2018})}\BibitemShut {NoStop}%
\bibitem [{\citenamefont {Ob{\v{s}}il}\ \emph
  {et~al.}(2018{\natexlab{a}})\citenamefont {Ob{\v{s}}il}, \citenamefont
  {Lachman}, \citenamefont {Pham}, \citenamefont {Le{\v{s}}und{\'a}k},
  \citenamefont {Hucl}, \citenamefont {{\v{C}}{\'\i}{\v{z}}ek}, \citenamefont
  {Hrabina}, \citenamefont {{\v{C}}{\'\i}p}, \citenamefont {Slodi{\v{c}}ka},\
  and\ \citenamefont {Filip}}]{obvsil2018nonclassical}%
  \BibitemOpen
  \bibfield  {author} {\bibinfo {author} {\bibfnamefont {P.}~\bibnamefont
  {Ob{\v{s}}il}}, \bibinfo {author} {\bibfnamefont {L.}~\bibnamefont
  {Lachman}}, \bibinfo {author} {\bibfnamefont {T.}~\bibnamefont {Pham}},
  \bibinfo {author} {\bibfnamefont {A.}~\bibnamefont {Le{\v{s}}und{\'a}k}},
  \bibinfo {author} {\bibfnamefont {V.}~\bibnamefont {Hucl}}, \bibinfo {author}
  {\bibfnamefont {M.}~\bibnamefont {{\v{C}}{\'\i}{\v{z}}ek}}, \bibinfo {author}
  {\bibfnamefont {J.}~\bibnamefont {Hrabina}}, \bibinfo {author} {\bibfnamefont
  {O.}~\bibnamefont {{\v{C}}{\'\i}p}}, \bibinfo {author} {\bibfnamefont
  {L.}~\bibnamefont {Slodi{\v{c}}ka}}, \ and\ \bibinfo {author} {\bibfnamefont
  {R.}~\bibnamefont {Filip}},\ }\bibfield  {title} {\enquote {\bibinfo {title}
  {Nonclassical light from large ensembles of trapped ions},}\ }\href
  {https://journals.aps.org/prl/abstract/10.1103/PhysRevLett.120.253602}
  {\bibfield  {journal} {\bibinfo  {journal} {Physical Review Letters}\
  }\textbf {\bibinfo {volume} {120}},\ \bibinfo {pages} {253602} (\bibinfo
  {year} {2018}{\natexlab{a}})}\BibitemShut {NoStop}%
\bibitem [{\citenamefont {Monz}\ \emph {et~al.}(2011)\citenamefont {Monz},
  \citenamefont {Schindler}, \citenamefont {Barreiro}, \citenamefont {Chwalla},
  \citenamefont {Nigg}, \citenamefont {Coish}, \citenamefont {Harlander},
  \citenamefont {H{\"a}nsel}, \citenamefont {Hennrich},\ and\ \citenamefont
  {Blatt}}]{monz2014}%
  \BibitemOpen
  \bibfield  {author} {\bibinfo {author} {\bibfnamefont {T.}~\bibnamefont
  {Monz}}, \bibinfo {author} {\bibfnamefont {P.}~\bibnamefont {Schindler}},
  \bibinfo {author} {\bibfnamefont {J.~T.}\ \bibnamefont {Barreiro}}, \bibinfo
  {author} {\bibfnamefont {M.}~\bibnamefont {Chwalla}}, \bibinfo {author}
  {\bibfnamefont {D.}~\bibnamefont {Nigg}}, \bibinfo {author} {\bibfnamefont
  {W.~A.}\ \bibnamefont {Coish}}, \bibinfo {author} {\bibfnamefont
  {M.}~\bibnamefont {Harlander}}, \bibinfo {author} {\bibfnamefont
  {W.}~\bibnamefont {H{\"a}nsel}}, \bibinfo {author} {\bibfnamefont
  {M.}~\bibnamefont {Hennrich}}, \ and\ \bibinfo {author} {\bibfnamefont
  {R.}~\bibnamefont {Blatt}},\ }\bibfield  {title} {\enquote {\bibinfo {title}
  {14-qubit entanglement: Creation and coherence},}\ }\href
  {https://journals.aps.org/prl/abstract/10.1103/PhysRevLett.106.130506}
  {\bibfield  {journal} {\bibinfo  {journal} {Physical Review Letters}\
  }\textbf {\bibinfo {volume} {106}},\ \bibinfo {pages} {130506} (\bibinfo
  {year} {2011})}\BibitemShut {NoStop}%
\bibitem [{\citenamefont {Keller}\ \emph {et~al.}(2019)\citenamefont {Keller},
  \citenamefont {Burgermeister}, \citenamefont {Kalincev}, \citenamefont
  {Didier}, \citenamefont {Kulosa}, \citenamefont {Nordmann}, \citenamefont
  {Kiethe},\ and\ \citenamefont {Mehlst{\"a}ubler}}]{keller2019controlling}%
  \BibitemOpen
  \bibfield  {author} {\bibinfo {author} {\bibfnamefont {J.}~\bibnamefont
  {Keller}}, \bibinfo {author} {\bibfnamefont {T.}~\bibnamefont
  {Burgermeister}}, \bibinfo {author} {\bibfnamefont {D.}~\bibnamefont
  {Kalincev}}, \bibinfo {author} {\bibfnamefont {A.}~\bibnamefont {Didier}},
  \bibinfo {author} {\bibfnamefont {A.}~\bibnamefont {Kulosa}}, \bibinfo
  {author} {\bibfnamefont {T.}~\bibnamefont {Nordmann}}, \bibinfo {author}
  {\bibfnamefont {J.}~\bibnamefont {Kiethe}}, \ and\ \bibinfo {author}
  {\bibfnamefont {T.}~\bibnamefont {Mehlst{\"a}ubler}},\ }\bibfield  {title}
  {\enquote {\bibinfo {title} {Controlling systematic frequency uncertainties
  at the 10- 19 level in linear coulomb crystals},}\ }\href
  {https://journals.aps.org/pra/abstract/10.1103/PhysRevA.99.013405} {\bibfield
   {journal} {\bibinfo  {journal} {Physical Review A}\ }\textbf {\bibinfo
  {volume} {99}},\ \bibinfo {pages} {013405} (\bibinfo {year}
  {2019})}\BibitemShut {NoStop}%
\bibitem [{\citenamefont {Kiethe}\ \emph {et~al.}(2017)\citenamefont {Kiethe},
  \citenamefont {Nigmatullin}, \citenamefont {Kalincev}, \citenamefont
  {Schmirander},\ and\ \citenamefont {Mehlst{\"a}ubler}}]{kiethe2017probing}%
  \BibitemOpen
  \bibfield  {author} {\bibinfo {author} {\bibfnamefont {J.}~\bibnamefont
  {Kiethe}}, \bibinfo {author} {\bibfnamefont {R.}~\bibnamefont {Nigmatullin}},
  \bibinfo {author} {\bibfnamefont {D.}~\bibnamefont {Kalincev}}, \bibinfo
  {author} {\bibfnamefont {T.}~\bibnamefont {Schmirander}}, \ and\ \bibinfo
  {author} {\bibfnamefont {T.}~\bibnamefont {Mehlst{\"a}ubler}},\ }\bibfield
  {title} {\enquote {\bibinfo {title} {Probing nanofriction and aubry-type
  signatures in a finite self-organized system},}\ }\href
  {https://www.nature.com/articles/ncomms15364} {\bibfield  {journal} {\bibinfo
   {journal} {Nature Communications}\ }\textbf {\bibinfo {volume} {8}},\
  \bibinfo {pages} {15364} (\bibinfo {year} {2017})}\BibitemShut {NoStop}%
\bibitem [{\citenamefont {Maier}\ \emph {et~al.}(2019)\citenamefont {Maier},
  \citenamefont {Brydges}, \citenamefont {Jurcevic}, \citenamefont {Trautmann},
  \citenamefont {Hempel}, \citenamefont {Lanyon}, \citenamefont {Hauke},
  \citenamefont {Blatt},\ and\ \citenamefont {Roos}}]{maier2019environment}%
  \BibitemOpen
  \bibfield  {author} {\bibinfo {author} {\bibfnamefont {C.}~\bibnamefont
  {Maier}}, \bibinfo {author} {\bibfnamefont {T.}~\bibnamefont {Brydges}},
  \bibinfo {author} {\bibfnamefont {P.}~\bibnamefont {Jurcevic}}, \bibinfo
  {author} {\bibfnamefont {N.}~\bibnamefont {Trautmann}}, \bibinfo {author}
  {\bibfnamefont {C.}~\bibnamefont {Hempel}}, \bibinfo {author} {\bibfnamefont
  {B.~P.}\ \bibnamefont {Lanyon}}, \bibinfo {author} {\bibfnamefont
  {P.}~\bibnamefont {Hauke}}, \bibinfo {author} {\bibfnamefont
  {R.}~\bibnamefont {Blatt}}, \ and\ \bibinfo {author} {\bibfnamefont {C.~F.}\
  \bibnamefont {Roos}},\ }\bibfield  {title} {\enquote {\bibinfo {title}
  {Environment-assisted quantum transport in a 10-qubit network},}\ }\href
  {https://journals.aps.org/prl/abstract/10.1103/PhysRevLett.122.050501}
  {\bibfield  {journal} {\bibinfo  {journal} {Physical Review Letters}\
  }\textbf {\bibinfo {volume} {122}},\ \bibinfo {pages} {050501} (\bibinfo
  {year} {2019})}\BibitemShut {NoStop}%
\bibitem [{\citenamefont {Ulm}\ \emph {et~al.}(2013)\citenamefont {Ulm},
  \citenamefont {Ro{\ss}nagel}, \citenamefont {Jacob}, \citenamefont
  {Deg{\"u}nther}, \citenamefont {Dawkins}, \citenamefont {Poschinger},
  \citenamefont {Nigmatullin}, \citenamefont {Retzker}, \citenamefont {Plenio},
  \citenamefont {Schmidt-Kaler},\ and\ \citenamefont
  {Singer}}]{ulm2013observation}%
  \BibitemOpen
  \bibfield  {author} {\bibinfo {author} {\bibfnamefont {S.}~\bibnamefont
  {Ulm}}, \bibinfo {author} {\bibfnamefont {J.}~\bibnamefont {Ro{\ss}nagel}},
  \bibinfo {author} {\bibfnamefont {G.}~\bibnamefont {Jacob}}, \bibinfo
  {author} {\bibfnamefont {C.}~\bibnamefont {Deg{\"u}nther}}, \bibinfo {author}
  {\bibfnamefont {S.}~\bibnamefont {Dawkins}}, \bibinfo {author} {\bibfnamefont
  {U.}~\bibnamefont {Poschinger}}, \bibinfo {author} {\bibfnamefont
  {R.}~\bibnamefont {Nigmatullin}}, \bibinfo {author} {\bibfnamefont
  {A.}~\bibnamefont {Retzker}}, \bibinfo {author} {\bibfnamefont
  {M.}~\bibnamefont {Plenio}}, \bibinfo {author} {\bibfnamefont
  {F.}~\bibnamefont {Schmidt-Kaler}}, \ and\ \bibinfo {author} {\bibfnamefont
  {K.}~\bibnamefont {Singer}},\ }\bibfield  {title} {\enquote {\bibinfo {title}
  {Observation of the kibble--zurek scaling law for defect formation in ion
  crystals},}\ }\href {https://www.nature.com/articles/ncomms3290} {\bibfield
  {journal} {\bibinfo  {journal} {Nature communications}\ }\textbf {\bibinfo
  {volume} {4}},\ \bibinfo {pages} {2290} (\bibinfo {year} {2013})}\BibitemShut
  {NoStop}%
\bibitem [{\citenamefont {Bruzewicz}\ \emph {et~al.}(2016)\citenamefont
  {Bruzewicz}, \citenamefont {McConnell}, \citenamefont {Chiaverini},\ and\
  \citenamefont {Sage}}]{bruzewicz2016scalable}%
  \BibitemOpen
  \bibfield  {author} {\bibinfo {author} {\bibfnamefont {C.~D.}\ \bibnamefont
  {Bruzewicz}}, \bibinfo {author} {\bibfnamefont {R.}~\bibnamefont
  {McConnell}}, \bibinfo {author} {\bibfnamefont {J.}~\bibnamefont
  {Chiaverini}}, \ and\ \bibinfo {author} {\bibfnamefont {J.~M.}\ \bibnamefont
  {Sage}},\ }\bibfield  {title} {\enquote {\bibinfo {title} {Scalable loading
  of a two-dimensional trapped-ion array},}\ }\href
  {https://www.nature.com/articles/ncomms13005} {\bibfield  {journal} {\bibinfo
   {journal} {Nature Communications}\ }\textbf {\bibinfo {volume} {7}},\
  \bibinfo {pages} {13005} (\bibinfo {year} {2016})}\BibitemShut {NoStop}%
\bibitem [{\citenamefont {Nizamani}\ \emph {et~al.}(2013)\citenamefont
  {Nizamani}, \citenamefont {Rind}, \citenamefont {Shaikh}, \citenamefont
  {Moghal},\ and\ \citenamefont {Saleem}}]{nizamani2013versatile}%
  \BibitemOpen
  \bibfield  {author} {\bibinfo {author} {\bibfnamefont {A.~H.}\ \bibnamefont
  {Nizamani}}, \bibinfo {author} {\bibfnamefont {M.~A.}\ \bibnamefont {Rind}},
  \bibinfo {author} {\bibfnamefont {N.~M.}\ \bibnamefont {Shaikh}}, \bibinfo
  {author} {\bibfnamefont {A.~H.}\ \bibnamefont {Moghal}}, \ and\ \bibinfo
  {author} {\bibfnamefont {H.}~\bibnamefont {Saleem}},\ }\bibfield  {title}
  {\enquote {\bibinfo {title} {Versatile ultra high vacuum system for ion trap
  experiments: Design and implementation},}\ }\href
  {https://www.researchgate.net/profile/Hussain_Saleem/publication/273127648_Versatile_Ultra_High_Vacuum_System_for_ION_Trap_Experiments_Design_and_Implementation/links/54f8394c0cf210398e94ad7f/Versatile-Ultra-High-Vacuum-System-for-ION-Trap-Experiments-Design-and-Implementation.pdf}
  {\bibfield  {journal} {\bibinfo  {journal} {Intl. Journal of Advancements in
  Research \& Technology}\ }\textbf {\bibinfo {volume} {2}} (\bibinfo {year}
  {2013})}\BibitemShut {NoStop}%
\bibitem [{\citenamefont {Calder}\ and\ \citenamefont
  {Lewin}(1967)}]{calder1967reduction}%
  \BibitemOpen
  \bibfield  {author} {\bibinfo {author} {\bibfnamefont {R.}~\bibnamefont
  {Calder}}\ and\ \bibinfo {author} {\bibfnamefont {G.}~\bibnamefont {Lewin}},\
  }\bibfield  {title} {\enquote {\bibinfo {title} {Reduction of stainless-steel
  outgassing in ultra-high vacuum},}\ }\href
  {https://iopscience.iop.org/article/10.1088/0508-3443/18/10/313/meta}
  {\bibfield  {journal} {\bibinfo  {journal} {British Journal of Applied
  Physics}\ }\textbf {\bibinfo {volume} {18}},\ \bibinfo {pages} {1459}
  (\bibinfo {year} {1967})}\BibitemShut {NoStop}%
\bibitem [{\citenamefont {Carr}\ \emph {et~al.}(2009)\citenamefont {Carr},
  \citenamefont {DeMille}, \citenamefont {Krems},\ and\ \citenamefont
  {Ye}}]{carr2009cold}%
  \BibitemOpen
  \bibfield  {author} {\bibinfo {author} {\bibfnamefont {L.~D.}\ \bibnamefont
  {Carr}}, \bibinfo {author} {\bibfnamefont {D.}~\bibnamefont {DeMille}},
  \bibinfo {author} {\bibfnamefont {R.~V.}\ \bibnamefont {Krems}}, \ and\
  \bibinfo {author} {\bibfnamefont {J.}~\bibnamefont {Ye}},\ }\bibfield
  {title} {\enquote {\bibinfo {title} {Cold and ultracold molecules: science,
  technology and applications},}\ }\href
  {https://iopscience.iop.org/article/10.1088/1367-2630/11/5/055049/meta}
  {\bibfield  {journal} {\bibinfo  {journal} {New Journal of Physics}\ }\textbf
  {\bibinfo {volume} {11}},\ \bibinfo {pages} {055049} (\bibinfo {year}
  {2009})}\BibitemShut {NoStop}%
\bibitem [{\citenamefont {Hansen}\ \emph {et~al.}(2012)\citenamefont {Hansen},
  \citenamefont {S{\o}rensen}, \citenamefont {Staanum},\ and\ \citenamefont
  {Drewsen}}]{hansen2012single}%
  \BibitemOpen
  \bibfield  {author} {\bibinfo {author} {\bibfnamefont {A.~K.}\ \bibnamefont
  {Hansen}}, \bibinfo {author} {\bibfnamefont {M.~A.}\ \bibnamefont
  {S{\o}rensen}}, \bibinfo {author} {\bibfnamefont {P.~F.}\ \bibnamefont
  {Staanum}}, \ and\ \bibinfo {author} {\bibfnamefont {M.}~\bibnamefont
  {Drewsen}},\ }\bibfield  {title} {\enquote {\bibinfo {title} {Single-ion
  recycling reactions},}\ }\href
  {https://onlinelibrary.wiley.com/doi/full/10.1002/anie.201203550} {\bibfield
  {journal} {\bibinfo  {journal} {Angewandte Chemie International Edition}\
  }\textbf {\bibinfo {volume} {51}},\ \bibinfo {pages} {7960--7962} (\bibinfo
  {year} {2012})}\BibitemShut {NoStop}%
\bibitem [{\citenamefont {Kimura}\ \emph {et~al.}(2011)\citenamefont {Kimura},
  \citenamefont {Okada}, \citenamefont {Takayanagi}, \citenamefont {Wada},
  \citenamefont {Ohtani},\ and\ \citenamefont
  {Schuessler}}]{kimura2011sympathetic}%
  \BibitemOpen
  \bibfield  {author} {\bibinfo {author} {\bibfnamefont {N.}~\bibnamefont
  {Kimura}}, \bibinfo {author} {\bibfnamefont {K.}~\bibnamefont {Okada}},
  \bibinfo {author} {\bibfnamefont {T.}~\bibnamefont {Takayanagi}}, \bibinfo
  {author} {\bibfnamefont {M.}~\bibnamefont {Wada}}, \bibinfo {author}
  {\bibfnamefont {S.}~\bibnamefont {Ohtani}}, \ and\ \bibinfo {author}
  {\bibfnamefont {H.~A.}\ \bibnamefont {Schuessler}},\ }\bibfield  {title}
  {\enquote {\bibinfo {title} {Sympathetic crystallization of cah+ produced by
  a laser-induced reaction},}\ }\href
  {https://doi.org/10.1103/PhysRevA.83.033422} {\bibfield  {journal} {\bibinfo
  {journal} {Phys. Rev. A}\ }\textbf {\bibinfo {volume} {83}},\ \bibinfo
  {pages} {033422} (\bibinfo {year} {2011})}\BibitemShut {NoStop}%
\bibitem [{\citenamefont {Chou}\ \emph {et~al.}(2017)\citenamefont {Chou},
  \citenamefont {Kurz}, \citenamefont {Hume}, \citenamefont {Plessow},
  \citenamefont {Leibrandt},\ and\ \citenamefont
  {Leibfried}}]{chou2017preparation}%
  \BibitemOpen
  \bibfield  {author} {\bibinfo {author} {\bibfnamefont {C.-w.}\ \bibnamefont
  {Chou}}, \bibinfo {author} {\bibfnamefont {C.}~\bibnamefont {Kurz}}, \bibinfo
  {author} {\bibfnamefont {D.~B.}\ \bibnamefont {Hume}}, \bibinfo {author}
  {\bibfnamefont {P.~N.}\ \bibnamefont {Plessow}}, \bibinfo {author}
  {\bibfnamefont {D.~R.}\ \bibnamefont {Leibrandt}}, \ and\ \bibinfo {author}
  {\bibfnamefont {D.}~\bibnamefont {Leibfried}},\ }\bibfield  {title} {\enquote
  {\bibinfo {title} {Preparation and coherent manipulation of pure quantum
  states of a single molecular ion},}\ }\href
  {https://www.nature.com/articles/nature22338} {\bibfield  {journal} {\bibinfo
   {journal} {Nature}\ }\textbf {\bibinfo {volume} {545}},\ \bibinfo {pages}
  {203} (\bibinfo {year} {2017})}\BibitemShut {NoStop}%
\bibitem [{\citenamefont {Rugango}\ \emph {et~al.}(2016)\citenamefont
  {Rugango}, \citenamefont {Calvin}, \citenamefont {Janardan}, \citenamefont
  {Shu},\ and\ \citenamefont {Brown}}]{rugango2016vibronic}%
  \BibitemOpen
  \bibfield  {author} {\bibinfo {author} {\bibfnamefont {R.}~\bibnamefont
  {Rugango}}, \bibinfo {author} {\bibfnamefont {A.~T.}\ \bibnamefont {Calvin}},
  \bibinfo {author} {\bibfnamefont {S.}~\bibnamefont {Janardan}}, \bibinfo
  {author} {\bibfnamefont {G.}~\bibnamefont {Shu}}, \ and\ \bibinfo {author}
  {\bibfnamefont {K.~R.}\ \bibnamefont {Brown}},\ }\bibfield  {title} {\enquote
  {\bibinfo {title} {Vibronic spectroscopy of sympathetically cooled cah+},}\
  }\href {https://onlinelibrary.wiley.com/doi/full/10.1002/cphc.201600645}
  {\bibfield  {journal} {\bibinfo  {journal} {ChemPhysChem}\ }\textbf {\bibinfo
  {volume} {17}},\ \bibinfo {pages} {3764--3768} (\bibinfo {year}
  {2016})}\BibitemShut {NoStop}%
\bibitem [{\citenamefont {Staanum}\ \emph {et~al.}(2010)\citenamefont
  {Staanum}, \citenamefont {H{\o}jbjerre}, \citenamefont {Skyt}, \citenamefont
  {Hansen},\ and\ \citenamefont {Drewsen}}]{staanum2010rotational}%
  \BibitemOpen
  \bibfield  {author} {\bibinfo {author} {\bibfnamefont {P.~F.}\ \bibnamefont
  {Staanum}}, \bibinfo {author} {\bibfnamefont {K.}~\bibnamefont
  {H{\o}jbjerre}}, \bibinfo {author} {\bibfnamefont {P.~S.}\ \bibnamefont
  {Skyt}}, \bibinfo {author} {\bibfnamefont {A.~K.}\ \bibnamefont {Hansen}}, \
  and\ \bibinfo {author} {\bibfnamefont {M.}~\bibnamefont {Drewsen}},\
  }\bibfield  {title} {\enquote {\bibinfo {title} {Rotational laser cooling of
  vibrationally and translationally cold molecular ions},}\ }\href
  {https://www.nature.com/articles/nphys1604} {\bibfield  {journal} {\bibinfo
  {journal} {Nature Physics}\ }\textbf {\bibinfo {volume} {6}},\ \bibinfo
  {pages} {271} (\bibinfo {year} {2010})}\BibitemShut {NoStop}%
\bibitem [{\citenamefont {Schneider}\ \emph {et~al.}(2010)\citenamefont
  {Schneider}, \citenamefont {Roth}, \citenamefont {Duncker}, \citenamefont
  {Ernsting},\ and\ \citenamefont {Schiller}}]{schneider2010all}%
  \BibitemOpen
  \bibfield  {author} {\bibinfo {author} {\bibfnamefont {T.}~\bibnamefont
  {Schneider}}, \bibinfo {author} {\bibfnamefont {B.}~\bibnamefont {Roth}},
  \bibinfo {author} {\bibfnamefont {H.}~\bibnamefont {Duncker}}, \bibinfo
  {author} {\bibfnamefont {I.}~\bibnamefont {Ernsting}}, \ and\ \bibinfo
  {author} {\bibfnamefont {S.}~\bibnamefont {Schiller}},\ }\bibfield  {title}
  {\enquote {\bibinfo {title} {All-optical preparation of molecular ions in the
  rovibrational ground state},}\ }\href
  {https://www.nature.com/articles/nphys1605} {\bibfield  {journal} {\bibinfo
  {journal} {Nature Physics}\ }\textbf {\bibinfo {volume} {6}},\ \bibinfo
  {pages} {275} (\bibinfo {year} {2010})}\BibitemShut {NoStop}%
\bibitem [{\citenamefont {Schiller}\ and\ \citenamefont
  {Korobov}(2005)}]{schiller2005tests}%
  \BibitemOpen
  \bibfield  {author} {\bibinfo {author} {\bibfnamefont {S.}~\bibnamefont
  {Schiller}}\ and\ \bibinfo {author} {\bibfnamefont {V.}~\bibnamefont
  {Korobov}},\ }\bibfield  {title} {\enquote {\bibinfo {title} {Tests of time
  independence of the electron and nuclear masses with ultracold molecules},}\
  }\href {https://journals.aps.org/pra/abstract/10.1103/PhysRevA.71.032505}
  {\bibfield  {journal} {\bibinfo  {journal} {Physical Review A}\ }\textbf
  {\bibinfo {volume} {71}},\ \bibinfo {pages} {032505} (\bibinfo {year}
  {2005})}\BibitemShut {NoStop}%
\bibitem [{\citenamefont {Petitprez}\ \emph {et~al.}(1989)\citenamefont
  {Petitprez}, \citenamefont {Lemoine}, \citenamefont {Demuynck}, \citenamefont
  {Destombes},\ and\ \citenamefont {Macke}}]{petitprez1989infrared}%
  \BibitemOpen
  \bibfield  {author} {\bibinfo {author} {\bibfnamefont {D.}~\bibnamefont
  {Petitprez}}, \bibinfo {author} {\bibfnamefont {B.}~\bibnamefont {Lemoine}},
  \bibinfo {author} {\bibfnamefont {C.}~\bibnamefont {Demuynck}}, \bibinfo
  {author} {\bibfnamefont {J.}~\bibnamefont {Destombes}}, \ and\ \bibinfo
  {author} {\bibfnamefont {B.}~\bibnamefont {Macke}},\ }\bibfield  {title}
  {\enquote {\bibinfo {title} {Infrared diode laser spectroscopy of cah and cad
  (x 2$\sigma$+). determination of mass-independent parameters},}\ }\href
  {https://aip.scitation.org/doi/abs/10.1063/1.456783} {\bibfield  {journal}
  {\bibinfo  {journal} {The Journal of Chemical Physics}\ }\textbf {\bibinfo
  {volume} {91}},\ \bibinfo {pages} {4462--4467} (\bibinfo {year}
  {1989})}\BibitemShut {NoStop}%
\bibitem [{\citenamefont {Guggemos}(2017)}]{guggemos2017}%
  \BibitemOpen
  \bibfield  {author} {\bibinfo {author} {\bibfnamefont {M.}~\bibnamefont
  {Guggemos}},\ }\emph {\bibinfo {title} {Precision spectroscopy with trapped
  $^{40}$Ca$^{+}$ and $^{27}$Al$^{+}$ ions}},\ \href
  {https://quantumoptics.at/images/publications/dissertation/2017_guggemos_diss.pdf}
  {Ph.D. thesis},\ \bibinfo  {school} {University of Innsbruck} (\bibinfo
  {year} {2017})\BibitemShut {NoStop}%
\bibitem [{\citenamefont {Kagwade}\ \emph {et~al.}(2001)\citenamefont
  {Kagwade}, \citenamefont {Clayton}, \citenamefont {Chidambaram},\ and\
  \citenamefont {Halada}}]{kagwade2001photochemical}%
  \BibitemOpen
  \bibfield  {author} {\bibinfo {author} {\bibfnamefont {S.~V.}\ \bibnamefont
  {Kagwade}}, \bibinfo {author} {\bibfnamefont {C.~R.}\ \bibnamefont
  {Clayton}}, \bibinfo {author} {\bibfnamefont {D.}~\bibnamefont
  {Chidambaram}}, \ and\ \bibinfo {author} {\bibfnamefont {G.~P.}\ \bibnamefont
  {Halada}},\ }\bibfield  {title} {\enquote {\bibinfo {title} {Photochemical
  breakdown of acetone on copper},}\ }\href
  {https://www.sciencedirect.com/science/article/pii/S0013468601003590}
  {\bibfield  {journal} {\bibinfo  {journal} {Electrochimica Acta}\ }\textbf
  {\bibinfo {volume} {46}},\ \bibinfo {pages} {2337--2342} (\bibinfo {year}
  {2001})}\BibitemShut {NoStop}%
\bibitem [{\citenamefont {Chambers}(2004)}]{chambers2004modern}%
  \BibitemOpen
  \bibfield  {author} {\bibinfo {author} {\bibfnamefont {A.}~\bibnamefont
  {Chambers}},\ }\href
  {https://books.google.cz/books?hl=en&lr=&id=AnvMBQAAQBAJ&oi=fnd&pg=PP1&dq=Modern+Vacuum+Physics+austin&ots=KQk-u2gPdt&sig=AZ96kSd9oiyWFT4ywWqrjMm8zNA&redir_esc=y#v=onepage&q=Modern%20Vacuum%20Physics%20austin&f=false}
  {\emph {\bibinfo {title} {Modern vacuum physics}}}\ (\bibinfo  {publisher}
  {CRC Press},\ \bibinfo {year} {2004})\BibitemShut {NoStop}%
\bibitem [{\citenamefont {Westerberg}\ \emph {et~al.}(1997)\citenamefont
  {Westerberg}, \citenamefont {Hj{\"o}rvarsson}, \citenamefont {Wall{\'e}n},\
  and\ \citenamefont {Mathewson}}]{westerberg1997hydrogen}%
  \BibitemOpen
  \bibfield  {author} {\bibinfo {author} {\bibfnamefont {L.}~\bibnamefont
  {Westerberg}}, \bibinfo {author} {\bibfnamefont {B.}~\bibnamefont
  {Hj{\"o}rvarsson}}, \bibinfo {author} {\bibfnamefont {E.}~\bibnamefont
  {Wall{\'e}n}}, \ and\ \bibinfo {author} {\bibfnamefont {A.}~\bibnamefont
  {Mathewson}},\ }\bibfield  {title} {\enquote {\bibinfo {title} {Hydrogen
  content and outgassing of air-baked and vacuum-fired stainless steel},}\
  }\href {https://www.sciencedirect.com/science/article/pii/S0042207X97000420}
  {\bibfield  {journal} {\bibinfo  {journal} {Vacuum}\ }\textbf {\bibinfo
  {volume} {48}},\ \bibinfo {pages} {771--773} (\bibinfo {year}
  {1997})}\BibitemShut {NoStop}%
\bibitem [{\citenamefont {Prins}\ and\ \citenamefont
  {Hermans}(1959)}]{prins1959}%
  \BibitemOpen
  \bibfield  {author} {\bibinfo {author} {\bibfnamefont {W.}~\bibnamefont
  {Prins}}\ and\ \bibinfo {author} {\bibfnamefont {J.~J.}\ \bibnamefont
  {Hermans}},\ }\bibfield  {title} {\enquote {\bibinfo {title} {Theory of
  permeation through metal coated polymer films},}\ }\href
  {https://pubs.acs.org/doi/pdf/10.1021/j150575a017} {\bibfield  {journal}
  {\bibinfo  {journal} {J. Phys. Chem.}\ }\textbf {\bibinfo {volume} {63}}
  (\bibinfo {year} {1959})}\BibitemShut {NoStop}%
\bibitem [{\citenamefont {Hankin}\ \emph {et~al.}(2019)\citenamefont {Hankin},
  \citenamefont {Clements}, \citenamefont {Huang}, \citenamefont {Brewer},
  \citenamefont {Chen}, \citenamefont {Chou}, \citenamefont {Hume},\ and\
  \citenamefont {Leibrandt}}]{hankin2019systematic}%
  \BibitemOpen
  \bibfield  {author} {\bibinfo {author} {\bibfnamefont {A.}~\bibnamefont
  {Hankin}}, \bibinfo {author} {\bibfnamefont {E.}~\bibnamefont {Clements}},
  \bibinfo {author} {\bibfnamefont {Y.}~\bibnamefont {Huang}}, \bibinfo
  {author} {\bibfnamefont {S.}~\bibnamefont {Brewer}}, \bibinfo {author}
  {\bibfnamefont {J.-S.}\ \bibnamefont {Chen}}, \bibinfo {author}
  {\bibfnamefont {C.}~\bibnamefont {Chou}}, \bibinfo {author} {\bibfnamefont
  {D.}~\bibnamefont {Hume}}, \ and\ \bibinfo {author} {\bibfnamefont
  {D.}~\bibnamefont {Leibrandt}},\ }\bibfield  {title} {\enquote {\bibinfo
  {title} {Systematic uncertainty due to background-gas collisions in
  trapped-ion optical clocks},}\ }\href {https://arxiv.org/abs/1902.08701}
  {\bibfield  {journal} {\bibinfo  {journal} {arXiv preprint arXiv:1902.08701}\
  } (\bibinfo {year} {2019})}\BibitemShut {NoStop}%
\bibitem [{\citenamefont {Hempel}(2014)}]{hempel2014digital}%
  \BibitemOpen
  \bibfield  {author} {\bibinfo {author} {\bibfnamefont {C.}~\bibnamefont
  {Hempel}},\ }\emph {\bibinfo {title} {Digital quantum simulation,
  Schr{\"o}dinger cat state spectroscopy and setting up a linear ion trap}},\
  \href {https://inis.iaea.org/search/search.aspx?orig_q=RN:47078884} {Ph.D.
  thesis},\ \bibinfo  {school} {University of Innsbruck} (\bibinfo {year}
  {2014})\BibitemShut {NoStop}%
\bibitem [{\citenamefont {Ob{\v{s}}il}\ \emph
  {et~al.}(2018{\natexlab{b}})\citenamefont {Ob{\v{s}}il}, \citenamefont
  {Le{\v{s}}und{\'a}k}, \citenamefont {Pham}, \citenamefont {Araneda},
  \citenamefont {{\v{C}}{\'\i}p}, \citenamefont {Filip},\ and\ \citenamefont
  {Slodi{\v{c}}ka}}]{obvsil2018scalable}%
  \BibitemOpen
  \bibfield  {author} {\bibinfo {author} {\bibfnamefont {P.}~\bibnamefont
  {Ob{\v{s}}il}}, \bibinfo {author} {\bibfnamefont {A.}~\bibnamefont
  {Le{\v{s}}und{\'a}k}}, \bibinfo {author} {\bibfnamefont {T.}~\bibnamefont
  {Pham}}, \bibinfo {author} {\bibfnamefont {G.}~\bibnamefont {Araneda}},
  \bibinfo {author} {\bibfnamefont {O.}~\bibnamefont {{\v{C}}{\'\i}p}},
  \bibinfo {author} {\bibfnamefont {R.}~\bibnamefont {Filip}}, \ and\ \bibinfo
  {author} {\bibfnamefont {L.}~\bibnamefont {Slodi{\v{c}}ka}},\ }\bibfield
  {title} {\enquote {\bibinfo {title} {Scalable phase interference from trapped
  ion chains},}\ }\href {https://arxiv.org/abs/1804.01518} {\bibfield
  {journal} {\bibinfo  {journal} {arXiv preprint arXiv:1804.01518}\ } (\bibinfo
  {year} {2018}{\natexlab{b}})}\BibitemShut {NoStop}%
\bibitem [{\citenamefont {F{\"u}rst}(2019)}]{furst2019trapped}%
  \BibitemOpen
  \bibfield  {author} {\bibinfo {author} {\bibfnamefont {H.~A.}\ \bibnamefont
  {F{\"u}rst}},\ }\emph {\bibinfo {title} {Trapped ions in a bath of ultracold
  atoms}},\ \href
  {https://dare.uva.nl/search?identifier=b6294950-12e7-4b9f-a4b8-fd73df09707a}
  {Ph.D. thesis},\ \bibinfo  {school} {Van der Waals-Zeeman Institute}
  (\bibinfo {year} {2019})\BibitemShut {NoStop}%
\end{thebibliography}

%

\end{document}